\documentclass{aa}
\usepackage{graphicx} 

\def\etal{{\sl et al.}\/ }
\def\kms{\,km~s$^{-1}$}
\def\micron{{\,$\mu$m}}
\def\SgrA{SgrA$^{\star}$}
\def\Paa{Pa$\alpha$}

\begin{document}

\title{The nature of the Galactic Center source IRS~13 revealed by high
spatial resolution in the infrared \thanks{This paper is based on
observations obtained with the Adaptive Optics System Hokupa'a/Quirc,
developed and operated by the University of Hawaii Adaptive Optics Group,
with support from the National Science Foundation,} \thanks{on observations
obtained with the Canada-France-Hawaii Telescope, operated by the National
Research Council of Canada, le Centre National de la Recherche Scientifique
de France and the University of Hawaii.}  }

\author{J.P.~Maillard\inst{1} \and T.~Paumard\inst{1} \and
S.R.~Stolovy\inst{2} \and F.~Rigaut\inst{3}
}
\authorrunning{Maillard \etal }
\titlerunning{GC source IRS~13 at high resolution}

\offprints{J.P.~Maillard} \mail{maillard@iap.fr}
\institute{Institut d'Astrophysique de Paris (CNRS), 98b Blvd Arago, 
75014 Paris, France 
\and  SIRTF Science Center, CalTech, MS 220-6, Pasadena, CA 91125, USA
\and Gemini North Headquarter, Hilo, HI 96720, USA
}

 \date{Received 30 July 2003 / Accepted 08 April 2004}  


\abstract{High spatial resolution observations in the 1 to 3.5\micron\/
region of the Galactic Center source known historically as \object{IRS~13}
are presented. They include ground-based adaptive optics images in the H,
Kp (2.12/0.4\micron) and L bands, HST-NICMOS data in filters between 1.1
and 2.2\micron, and integral field spectroscopic data from BEAR, an Imaging
FTS, in the \ion{He}{i} 2.06\micron\/ and the Br$\gamma$ line
regions. Analysis of all these data provides a completely new picture of
the main component, \object{IRS~13E}, which appears as a cluster of seven
individual stars within a projected diameter of $\sim$\,0.5\arcsec\ (0.02
pc). The brightest sources, 13E1, 13E2, 13E3 which is detected as a binary,
and 13E4, are all massive stars of different type. The star 13E1 is a
luminous, blue object, with no detected emission line. 13E2 and 13E4 are
two hot, high-mass emission line stars, 13E2 being at the WR stage and 13E4
a massive O-type star. In contrast, 13E3A and B are extremely red objects,
proposed as other examples of dusty WR stars, like \object{IRS~21} (Tanner
\etal \cite{tanner}). All these sources have a common westward proper
motion (Ott \etal \cite{ott2}) indicating they are bounded.  Two other
sources, detected after deconvolution of the AO images in the H and Kp
bands, are also identified. One, that we call 13E5, is a red source similar
to 13E3A and B, while the other one, 13E6, is probably a main sequence
O~star in front of the cluster.  Considering this exceptional concentration
of comoving massive hot stars, IRS~13E is proposed as the remaining core of
a massive star cluster, which could harbor an intermediate-mass black hole
(IMBH) (Portegies Zwart \& McMillan \cite{zwart2}) of
$\sim$\,1300\,M$_{\odot}$. This detection plays in favor of a scenario,
first suggested by Gerhard (\cite{gerhard}), in which the helium stars and
the other hot stars in the central parsec originate from the stripping of a
massive cluster formed several tens of pc from the center. This cluster
would have spiraled towards \SgrA, and IRS~13E would be its
remnant. Furthermore, IRS~13E might be the second black hole needed
according to a model by Hansen \& Milosavljevi\'c (\cite{hansen}) to drag
massive main-sequence stars, in the required timescale, very close to the
massive black hole.  The detection of a discrete X-ray emission (Baganoff
\etal \cite{baganoff}) at the IRS~13 position (within the positional
accuracy) is examined in this context.  }

\maketitle

\section{Introduction}

In the early images of the central region of the GC recorded in the near
infrared, at the best seeing-limited resolution, several bright point
sources dominate the $\sim$\,20\arcsec$\times$20\arcsec\/ field centered on
\SgrA. With the radical improvement of angular resolution through multiple
short exposures with shift-and-add (SHARP camera, Eckart \etal
\cite{eckart}) or speckle techniques (Ghez \etal \cite{ghez}), with the
images from the NICMOS cameras on board HST (Stolovy \etal \cite{stolovy}),
to the advent of adaptive optics correctors behind large telescopes as NAOS
(Ott \etal \cite{ott3}), a more complex vision of this crowded field has
emerged. At a spatial resolution in the best case of 0.05\arcsec\/ (Ghez
\etal \cite{ghez}) new, faint stellar sources appear and the
early-identified sources are often resolved into several
components. Therefore, the observed spectra, generally made at a resolution
limited at best by a slit not less than 1\arcsec\/ wide, are actually
composites of emission from stellar objects of different spectral type as
well as from emission lines from the surrounding interstellar medium (ISM).
The stellar type identified from these spectra can be wrong if attributed
to a single source, which can erroneously look unusual. This care led to
the paper on the revised identifications of the central cluster of
\ion{He}{i} stars by Paumard \etal (\cite{paumard}) made from slitless
integral field spectroscopy, with BEAR an imaging Fourier Transform
Spectrometer (Maillard \cite{maillard}). This instrument makes it possible
to obtain near-infrared spectroscopy at the seeing-limited resolution of
Mauna Kea (i.e. $\simeq$\,0.5\arcsec), which represents a significant
improvement.  However, this resolution is not sufficient for such a crowded
field. Therefore, the data were compared to adaptive optics (AO) images in
the K band of the same field, with a spatial resolution of
$\sim$\,0.15\arcsec, in order to check whether the emission sources at the
angular resolution of seeing were single stars or not.

Among the sources studied in this paper the object historically named
IRS~13, is a typical example. Located approximately 3.6\arcsec\/ south-west
of \SgrA, at the edge of the circular structure of the \object{SgrA~West}
HII region called the Minicavity, it appears in early works at
near-infrared wavelengths, in J, H, K (Rieke \etal \cite{rieke}) and L
(Allen \& Sanders \cite{allen}), as a bright spot. However, from a
discussion on the sources of energy at the Galactic Center, Rieke \etal
(\cite{rieke}) note \lq\lq there are separate luminosity sources in the
core of the 10\micron\/ sources 1, 9 and 13\rq\rq\/, because, particularly
for IRS~13, the energy distribution from 1 to 5\micron\/ presents a sudden,
steep increase beyond 3\micron. A first work at subarcsecond resolution, by
lunar occultation in the K band (Simon \etal \cite{simon}), indicates that
IRS~13 resolves into a pair of sources, separated by $\sim$\,1.2\arcsec,
which were henceforth designated as IRS~13E and 13W.  In the photometric
survey of Ott \etal (\cite{ott}) from SHARP imaging data (Eckart \etal
\cite{eckart}), at a limit of resolution of $\sim$\,0.15\arcsec\/ obtained
by deconvolution, two equally bright components (K$_{mag}$\,=\,10.26) are
reported, IRS~13E1 and 13E2, with a separation of 0.2\arcsec.  Paumard
\etal \cite{paumard}) published the first AO map of IRS~13E in the K band,
extracted from a larger image of the central region obtained with the
CFHT-AO system (Lai \etal \cite{lai}). We noted that a fainter third
source, we called \object{IRS~13E3}, was also present, forming a kind of
equilateral triangle with 13E1 and 13E2. In parallel, Cl\'enet \etal
(\cite{clenet}) published a photometric analysis of the same data.  For
IRS~13E they reported three components in the K band, noted 13E1, 13E2 and
13N. From the given offset coordinates the source called 13N in this paper
is not exactly coincident with the source we had previously called 13E3
(Paumard \etal \cite{paumard}).

In the meantime, spectroscopic works on the stellar population of the
central parsec were conducted. Several spectra of IRS~13 (Blum \etal
\cite{blum95}, Libonate \etal \cite{libo}, Tamblyn \etal \cite{tamblyn}),
and specifically of IRS~13E ( Krabbe \etal \cite{krabbe}, Genzel \etal
\cite{genzel96}, Najarro \etal \cite{najarro}), and of \object{IRS~13W}
(Krabbe \etal \cite{krabbe}) have been published. They cover mostly the K
band, i.e. all or part of the 1.95 -- 2.45\micron\/ range, and one the 1.57
-- 1.75\micron\/ region of the H band (Libonate \etal \cite{libo}).  From
these spectra IRS~13W is unambiguously a cool star with the strong vib-rot
CO signatures at 2.3\micron.  Emission lines were detected on IRS~13E,
typical of a luminous, hot star: strong \ion{He}{i} 2.058, 2.112\micron,
Br$\gamma$ line and other Brackett lines up to Br12, plus weaker lines of
[\ion{Fe}{ii}], [\ion{Fe}{iii}], and a weak emission at 2.189\micron\/
attributed to \ion{He}{ii}.  The 7--10 line is the most intense of this ion
in the K band, and is supposed unblended, while the 8--14 line cannot be
distinguished from Br$\gamma$ of which it is separated by less than
2\,cm$^{-1}$. The \ion{He}{ii} line if detected is a precious type
indicator in the classification of WR stars (Tamblyn \etal \cite{tamblyn},
Figer \etal \cite{figer97}).  From these spectral characteristics Libonate
\etal (\cite{libo}) concluded \lq\lq the IRS~13 spectrum bears a strong
resemblance to the low-resolution K-band spectra of P~Cygni (a LBV) and the
AF source\rq\rq\/. Krabbe \etal (\cite{krabbe}) proposed IRS~13E as a WN9
star.  From a K-band spectrum of the Minicavity Lutz \etal (\cite{lutz})
attributed to [\ion{Fe}{iii}] several emission lines (2.145, 2.218, 2.242,
2.348\micron). Images at 2.218\micron\/ show that this line is present all
over the Minicavity, with a particular enhancement at the position of
IRS~13.  From NICMOS data in the F164N filter (Stolovy \etal
\cite{stolovy}) the lower ionized Fe species, [\ion{Fe}{ii}] is detected in
emission in the central ISM, particularly strong at the edge of the
Minicavity, but remarkably absent at the position of IRS~13E. As mentioned
in this paper the ionization condition for [\ion{Fe}{iii}] requires
16.2\,eV while only 7.2\,eV are required for [\ion{Fe}{ii}]. Hard
ionization radiation is likely originating from IRS~13E. Hence, it can
concluded that these iron lines are not of stellar origin but are present
in the surrounding ionized gas.

From the extraction of the \ion{He}{i} 2.058\micron\/ line profile at high
spectral resolution from the BEAR data, Paumard \etal (\cite{paumard})
concluded that one of the three sources identified as forming IRS~13E,
instead of a LBV-type star should be already at the WR stage. The main
argument was the width of the observed emission line profile (FWHM
$\simeq$\,974\kms), typical of WR stars, making the source belonging to a
class of broad-line stars including 8 other stars in the central cluster of
helium stars.  It was also measured that the broad-line stars were in
average weaker by $\simeq$\,2.4 mag in the K band than the rest of helium
sources which were characterized by a narrow \ion{He}{i} emission
line. With a K$_{mag}$ for IRS~13E3 consistent with the magnitude measured
for the other broad-line stars, and 13E1 and 13E2 much brighter, it was
proposed that IRS~13E3 should be the WR-type helium emitter. of because and
It was not possible to precise the 13E1 and 13E2 stellar type.  

 In the centimetric domain, Zhao \& Goss (\cite{zhao}) presented the
 detection of IRS~13 at 7 and 13~mm with the VLA, at a resolution
 of 0.06\arcsec, which they reported to be the brightest radio continuum
 source after \SgrA\/ at the Galactic Center.  They resolved the source
 into two components, one with no significant proper motion while the other
 one is moving south at a rate 6.2\,$\pm$\,1.1\,mas~yr$^{-1}$. They called
  the two compact \ion{H}{ii} regions IRS~13E and IRS~13W, which
 was improper, since these denominations had already been given to infrared
 sources at IRS~13 as reminded above, with which they are not
 coincident. However, this detection is another element which makes this
 source special.

The detection of a bright, discrete X-ray emission source (Baganoff \etal
\cite{baganoff}, Muno \etal \cite{muno}) at the IRS~13 position
within the positional accuracy -- source \object{CXOGC J174539.7-290029} -- among the
brightest sources within the central parsec besides \SgrA, is a last
element contributing to make this source an object of interest.  This coincidence
already triggered the interest of Coker \etal (\cite{coker}) who presented
the first Chandra X-ray spectrum of IRS~13 and deduced that it was
consistent with a highly absorbed X-ray binary system. They concluded that
IRS~13E2 should be a compact post-LBV binary whose colliding winds were
the source of the X-ray emission. 

As a conclusion, the origin of the brightness of IRS~13, one of the
brightest objects at all wavelengths in the vicinity of \SgrA, has been a
matter of debate.  We examined all the high-angular resolution images in
the near infrared currently available.  This analysis made it possible to
build a completely new picture of IRS~13 which is presented in this
paper. We will show that the peculiar spectral energy distribution (SED)
previously reported is well explained by the nature of the individual
sources which compose IRS~13. Unexpectedly, this analysis provides new
insight to the unsolved question of the formation of the central massive
cluster. It has also given the opportunity through the study of the stars
in the IRS~13 field to characterize a sample of the star population in the
central parsec.  A preliminary version of the paper was presented at the
Galactic Center Workshop 2002 (Maillard \etal \cite{maillard2}).
\section{High-angular resolution images of IRS~13 and data reduction}

Ground-based AO images from several telescopes and space-based NICMOS data
containing IRS~13 in their field have been gathered. Their
characteristics are given in Table~\ref{instrum}.

        \subsection{Adaptive Optics data} 

\begin{figure}[!ht]
\begin{center}
\resizebox{\hsize}{!}{\includegraphics{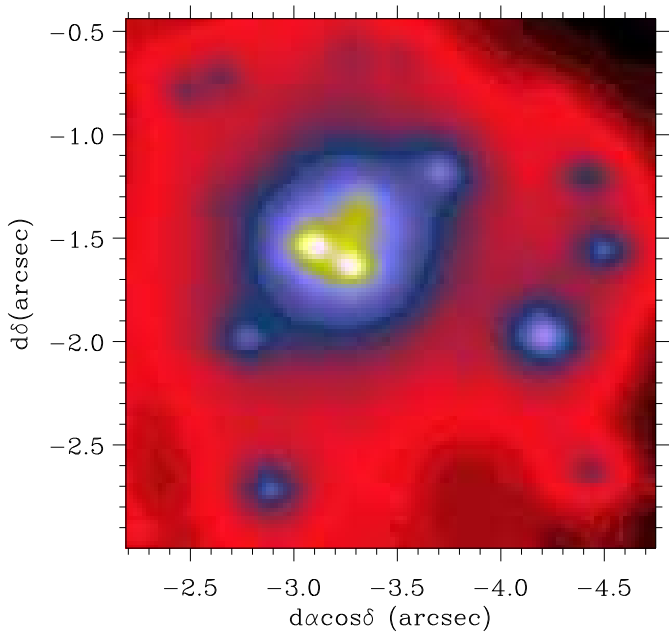}} 
\caption[]{IRS~13 field from the Gemini AO image in the Kp band. IRS~13E is the central,
compact group of stars and IRS~13W the brightest source
$\sim$\,1\arcsec\/ southwest of IRS~13E. The coordinates are in arcsec
offset from \SgrA.}

         \label{Kp_AO_f1}
\end{center}
   \end{figure}

We have analyzed data from three different AO systems: PUEO/CFHT (Lai \etal
\cite{lai}), Hokupa'a/Gemini North (Graves \etal \cite{graves}), and
Adonis/ESO 3.6-m telescope (Beuzit \etal \cite{beuzit}).  The CFHT data, in
the K band, were obtained on 26 June, 1998. They were already presented and
analyzed in detail in Paumard \etal (\cite{paumard}). The total field
covers 40\arcsec$\times$40\arcsec\/ centered on \SgrA. The FWHM of the
Point Spread Function (PSF) varies from 0.13\arcsec\/ to 0.20\arcsec\/ in
the field since PUEO has a visible wavefront sensor, requiring to use a
V~$\simeq$\,14 star 20\arcsec\ to the north-east of the field center. The
L-band data were obtained with the ADONIS visible wavefront sensor in 2000,
from 20 to 22 May, on the ESO 3.6-m telescope. These data are described in
Cl\'enet \etal (\cite{clenet}). The final L-band image we used has a PSF
with a FWHM of 0.291\arcsec\/ and covers a field of
$\simeq$\,13\arcsec$\times$13\arcsec\/ centered on \SgrA. The Gemini North
data were part of the AO demonstration run conducted by F.~Rigaut on the
Galactic Center in July 2000. The data were obtained with the Hokupa'a AO
system and the QUIRC camera (Graves \etal \cite{graves}) in the Kp and the
H band (Table~\ref{instrum}) respectively 3 and 6 July in field 1, centered
on \SgrA.  The field coverage of each image is 20\arcsec$\times$20\arcsec.
For the H image the FWHM of the PSF varies from 0.115 to 0.19\arcsec\/ and
for the Kp image from 0.12 to 0.18\arcsec. In the vicinity of IRS~13 the
measured FWHMs are respectively 0.180\arcsec\/ and 0.172\arcsec. These AO
data are characterized by a low Strehl ratio of 2.5$\%$ in H and 7$\%$ in
Kp.  The small portion of the AO image in the Kp filter analyzed in the
paper is shown on Fig.~\ref{Kp_AO_f1}.

        \subsection{NICMOS data} 

Six filters, coded F110M, F145M, F160W, F187N, F190N, and F222M were used
in observing the stars in the inner parsec of the Galactic Center with the
NICMOS cameras on board HST, as part of three independent Galactic Center
programs (7214, 7225, and 7842). These data were taken between Aug. 1997
and Aug. 1998. The filter properties (central wavelength, bandwidth,
zero-magnitude flux, pixel size) and the data processing of the raw data
are described in Rieke (\cite{riekem}) and Stolovy \etal (\cite{stolovy}).
Although each program covered a larger region, we extracted a small portion
of $\simeq$\,4\arcsec$\times$4\arcsec\/ centered on IRS~13E of the image
from each filter. All these diffraction-limited images were particularly
useful to derive the SED of the stellar components of IRS~13 and its
environment.

Apart from wide (W) or medium (M) bandpass filters, F187N is a 1$\%$
narrow-band filter centered on the \Paa\/ line.  By subtraction of F190N, a
comparable narrow-band filter in the nearby continuum, the distribution of
the ionized gas in the inner parsecs was obtained (Stolovy \etal
\cite{stolovy}). In this map are also discrete emission sources which must
come from \Paa\/ emission in the atmosphere of hot stars.  We used a
best-fit scale factor between F187N and F190N in order to minimize both
positive and negative stellar residuals of the stars in the central parsec.
Negative residuals can arise in stars with \Paa\/ absorption and/or in
stars with local excess extinction along the line of sight. For this paper,
we identified emission line stars as those with very significant
F187N/F190N ratios (exceeding 1.2).  The portion of the resulting \Paa\/
image with the IRS~13 complex was essential in determining which of the
individual stars was an emission line star (Fig.~\ref{Paa}).

\begin{table}[!ht]
\caption {Instruments, filter properties and year of data acquisition of
the high-resolution images of the IRS~13 field}

\begin{center}
  \begin{tabular}{lc@{}cl}
   \hline\hline
   Filter & Instrument / Telescope & $\lambda$/{\bf $\Delta\lambda$} & year\\
          &                        &   ($\mu$m)       & \\
   \hline
   F110M &      NICMOS/HST            & 1.10/0.200 &97-98 \\
   F145M &      NICMOS/HST            & 1.45/0.197 &97-98\\
   F160W &      NICMOS/HST            & 1.60/0.400 &97-98\\
   H     &      Hokupa'a+Quirc/Gemini N & 1.65/0.296 &2000\\
   Pa$\alpha$ & NICMOS/HST            & 1.87/0.019 &97-98\\
   F190N &      NICMOS/HST            & 1.90/0.017 &97-98\\
   Kp    &      Hokupa'a+Quirc/Gemini N  & 2.12/0.410 & 2000 \\
   K     &      PUEO+KIR/CFHT         & 2.20/0.340 &1998\\
   F222M &      NICMOS/HST            & 2.22/0.143 &97-98 \\
   L     &      ADONIS+COMIC/3.6m~ESO~ & 3.48/0.590 & 2000\\
   \hline
  \end{tabular}
\end{center}
\label{instrum}
\end{table}
\vspace{-0.5truecm}

\subsection{Data reduction}
\label{reduc}
The initial processing of each dataset is not described here. It can be
   found in the papers cited with the presentation of each of them. Below
   are listed the main operations which were performed subsequently to the
   initial data reduction.

   \subsubsection{Methods of star extraction}

 To perform the extraction of the individual stars from the AO and NICMOS
 data, $StarFinder$, an IDL procedure (Diolaiti \etal \cite{diolaiti})
 specially written for AO data, was first used.  The preliminary operation
 consists of building a good PSF by averaging several isolated and bright
 stars in each image. By adjusting the PSF to the local peaks in the field
 the exact position and the flux of the corresponding stars are
 retrieved by the procedure.
   
	  The image residuals left by applying this procedure led us to
suspect that more sources might be present in the IRS~13 complex, which
could not be detected because the spatial resolution was not high
enough to separate them. To improve the star detection we applied a new
deconvolution code called MCS (Magain \etal \cite{mcs}).  Contrary to the
{\em StarFinder} procedure, the MCS program uses an analytical PSF (a
Moffat-type function). The final PSF is chosen prior to deconvolution, with
a width compatible with the original image sampling. Thus, the final PSF
can be narrower than the observed one without violating the Shannon
sampling theorem. The contribution of a continuous background is also
matched to the image with an adjustable smoothing.  The sampling was high
enough for the Gemini AO images, in the H and the Kp bands, to provide a
substantial gain in resolution. Because of the sampling of the L-band image
a more limited gain was obtained, which was useful
anyway. For the H and the Kp images the width of the synthetic PSF is equal
to 0.040\arcsec\/ and to 0.192\arcsec\/ for the L-band image, i.e. a gain
in resolution respectively of a factor 4.5 in H, 4.3 in Kp and 1.5 in
L. However, this method applied to the IRS~13 images demanded many trials
to reach a stable and plausible solution because of the presence of a non
uniform background in which the sources are embedded. The convergence was
helped by the comparison between the solutions for the H and Kp filters. We
imposed the sources to be detected in both bands, to avoid the risk of
taking deconvolution artifacts for sources.  The two filter bandpasses are
almost adjacent which makes pertinent the detection in both filters, even
for very red or very blue objects.  This code was not applied to the CFHT
K-band nor to the NICMOS images, which both PSFs show significant secondary
rings, because they are not correctly handled by this code, written for
seeing-limited images. Anyway, it was applied on the Gemini/Hokupa'a data
because of their low Strehl ratio.

   \subsubsection{Calibration} 

All the NICMOS data were calibrated as described by Rieke
 (\cite{riekem}).  The calibration of the sources in the L band was based
 on the photometry of IRS~13W reported by Cl\'enet \etal (\cite{clenet}), a
 prominent, isolated source in this band. The Gemini data in the H and the
 Kp filters were not calibrated.  In order to minimize photometric
 uncertainties between data from various origins we calibrated the two
 filters by interpolation from the NICMOS data which offer a set of filters
 close to H and Kp.  Based on IRS~13E1, a bright and well-measured star by
 MCS deconvolution in the stellar complex, and taking into account the
 exact filter bandpasses, we applied a strict linear interpolation between
 the F160W and F190N flux of this star for the H band (central wavelength
 1.65\micron), between F190N and F222M for the Kp band (central wavelength
 2.12\micron, see Table~\ref{instrum}). With this source calibrated that
 way in the H and the Kp filters the flux measurements of all the other
 sources in the two filters were calibrated. The intensities of the
 detected sources in the 8 filters were expressed in $\mu$Jy as were
 originally the NICMOS data.

   \subsubsection{Proper motions} 

We obtained from Ott \etal (\cite{ott2}), who have conducted an analysis of
ten years of SHARP data, providing more than 1000 proper motions in the
central parsec, the proper motions of the main components of IRS~13 and of
its nearby field.

\subsection{BEAR data}
Even though the BEAR data are not at the spatial resolution of the NICMOS
and AO data, the Br$\gamma$ and 2.06\micron\/ \ion{He}{i} line profile at
IRS~13E were used as complementary information to help determine the
spectral type of the underlying stars.  The IRS~13 complex is located in a
region of interstellar emission, intense in Br$\gamma$ (Morris \& Maillard
\cite{morris}, Paumard \etal \cite{paumard2}) as well as in the
2.06\micron\/ helium line (Paumard \etal \cite{paumard}). Hence, the line
profiles had to be corrected to remove the ISM contribution, made here of
two main velocity components (Fig.~\ref{HeI_Brg} - central panel),
superimposed to the stellar profile. This operation required special
attention for the Br$\gamma$ profile since the ISM emission is locally very
intense. Only the high resolution (21.3\kms) made it possible to separate,
with some approximation anyway, the stellar profile which is naturally
broader, from the interstellar components.
 
\section{Results}
From the 8 high-resolution images between 1.1 and 3.5\micron\/
(Table~\ref{instrum}) and the spectroscopic data available, the following
results on the IRS~13 complex and its environment have been obtained:

\subsection{Star detection}

  Twenty sources in total (Table~\ref{detec}) are detected both in H and Kp
after MCS deconvolution of the Gemini data in the small IRS~13 field, which
are identified on Fig.~\ref{Kp_dec_IRS}. All these sources were searched
for with $StarFinder$ for the filters where the deconvolution operation
could not be applied, and again with the MCS code in the L band.  Empty
positions in Table~\ref{detec} indicate that the source is not detectable
in a specific filter. An upper limit is given at some positions for the
extreme filters (1.1\micron\/ and L) where the detection is the most
difficult. This limit is not uniform in the field depending on the
proximity of another bright source, particularly in L.

\begin{table}[!ht]
\caption {Photometric measurements in all the filters, in log($\mu$Jy), of
all the stars detected by MCS deconvolution (Sect.~\ref{reduc}) of the
2.5\arcsec$\times$2.5\arcsec\/ IRS~13 field from the H and Kp Gemini AO
images.  Measurements in the NICMOS filters (F) were obtained by
$StarFinder$, and in the L band by MCS analysis. The sources are listed by
order of decreasing brightness in the Kp band, except IRS~13W, which is
listed first. Values preceded by $<$ in the 1.1\micron\/ and the L bands are
minima of detection, i.e. flux upper limits.}

\begin{center}
        \begin{tabular}{|l@{}|@{}r@{} c@{} c@{} c@{} c@{} | c | c@{} c |}
\hline
ID & F110M &~F145M&~F160W& ~H~& F190N & Kp & F222M & L\\
\hline\hline
W & $<$0.56 & 2.58 & 3.28 & 3.20 & 3.96 & 4.27 & 4.36 & 4.89\\
E1 & 1.76 & 3.40 & 3.82 & 3.93 & 4.27 & 4.40 & 4.49 & 5.03\\
E2 & 1.61 & 3.22 & 3.71 & 3.80 & 4.17 & 4.38 & 4.48 & 5.37\\
E4 & 0.93 & 2.68 & 3.17 & 3.28 & 3.69 & 4.11 & 4.18 & $<$4.66\\
5 & 0.78 & 2.66 & 3.20 & 3.30 & 3.72 & 4.04 & 3.93 &    \\
6 & 0.39 & 2.51 & 3.07 & 3.14 & 3.57 & 3.94 & 3.89 &    \\
7 & 1.23 & 2.78 & 3.20 & 3.26 & 3.63 & 3.92 & 3.88 &    \\
8 & 0.62 & 2.55 & 3.08 & 3.08 & 3.53 & 3.90 & 3.81 &    \\
E3A & $<$0.45 & & 2.29 & 2.43 & 3.11 & 3.81 &      & 5.46\\
10 & 0.11 & 1.88 & 2.63 & 2.81 & 3.23 & 3.68 & 3.55 &    \\
E3B & $<$0.45 &  & 2.11 & 2.24 & & 3.57 &    & 5.29\\
12 & 0.69 & 1.98 & 2.62 & 2.66 & 3.15 & 3.48 & 3.45 &    \\
13 & -0.09 & 2.02 & 2.58 & 2.65 & 3.14 & 3.46 & 3.40 &   \\
14 & -0.16 & 2.01 & 2.55 & 2.65 & 3.11 & 3.43 & 3.34 &    \\
15 & -0.06 & 1.75 & 2.25 & 2.37 & 2.82 & 3.12 & 3.23 &   \\
16 & & 1.57 & 2.18 & 2.36 & 2.71 & 3.11 & 2.92 &     \\
E5 & $<$-0.17 & 1.60 & & 1.75 & 2.72 & 3.08 & 3.61 & 5.07\\
18 & 0.56 & 1.87 & 2.27 & 2.30 & 2.74 & 2.99 & 3.02 &   \\
E6 & $<$0.48 & & 2.46 & 2.53 & & 2.97 &     & $<$4.68\\
20 & 0.86 & 1.42 & 2.01 & 2.11 & 2.57 & 2.88 & 2.89 &    \\
\hline
        \end{tabular}
\end{center}
  \label{detec}
\end{table}

In conclusion, IRS~13E is resolved into a compact cluster of at least 7
objects encircled within $\simeq$\,0.5\arcsec\/ (Fig.~\ref{Kp_dec_IRS}). The two
brightest sources, 13E1 and 13E2, had already been identified in Ott \etal
(\cite{ott}) and in Paumard \etal (\cite{paumard}). The one we had noted
13E3 appears double after deconvolution. Thus, we call the two components
13E3A and 13E3B. By continuity, the closest bright source, north of 13E3, is
called 13E4. Two other sources revealed by deconvolution are called 13E5
and 13E6. 

        \subsection{Astrometry}
\label{astrometry}
\begin{figure}[!ht]
\begin{center}
\resizebox{\hsize}{!}{\includegraphics{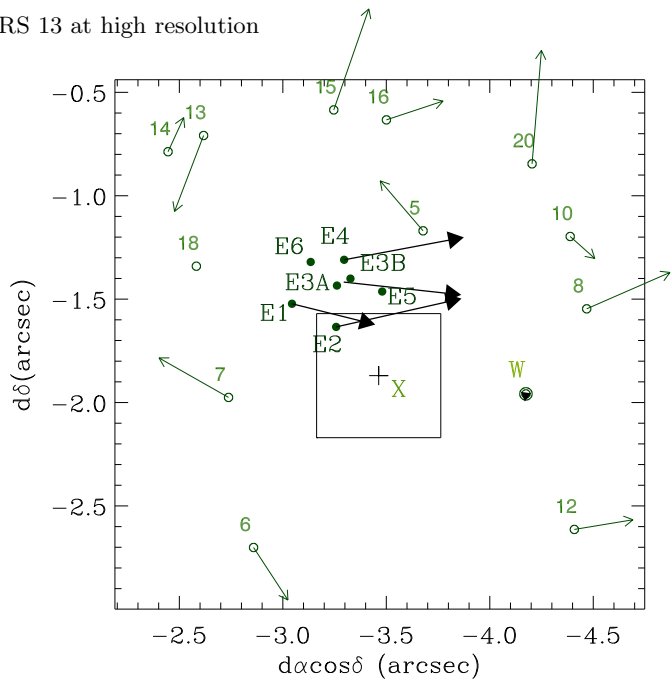}} \caption[]{The
 Fig.~\ref{Kp_AO_f1} field with the star detected after deconvolution by
 the MCS code (Sect.~\ref{reduc}). The vector associated with most of the
 stars represents in amplitude the velocity and the direction of proper
 motions measured from SHARP data by Ott \etal (\cite{ott2}).  For E3A and
 E3B, only the proper motion of the center of light is determined. The
 amplitudes reported in Table~\ref{mvprop} for the four brightest sources
 of IRS~13E give the scale of the proper motion vectors.  The cross marked
 X represents the nominal position of the X-ray source at the center of an
 error box of $\pm$\,0.3\arcsec (see Sect.~\ref{astrometry}) .}

   \label{Kp_dec_IRS}
\end{center}
   \end{figure}

 The positions of the sources in the 2.5\arcsec$\times$2.5\arcsec\/ IRS~13
 field in offsets from \SgrA\/ are given in Table~\ref{astrom}.  The
 astrometry was retrieved from the results of the deconvolution of the
 portion of the Gemini Kp-band image.  The precision of the positions
 depends on the brightness of the sources. For the brightest sources the
 relative 1-$\sigma$ position error is equal to 1.0\,mas.  From these
 positions the projected angular separations between the main IRS~13E
 components are given in Table~\ref{sepa}, translated to AU, by taking a
 distance to the GC of 8\,kpc (Reid 1993). All the sources of the small
 field are detected by the reprocessing of the SHARP data we obtained prior
 to publication from Ott \etal (\cite{ott2}), except that a single source
 is given for E3A and E3B ($\#$118) and E6 is not detected.  The
 Table~\ref{astrom} offsets are estimated from the positions of isolated
 stars around IRS~13E published by Ott \etal (\cite{ott}) which, in this
 paper are given relative to \object{IRS~7}. The offset of IRS~7 from
 \SgrA\/ estimated from VLA measurements, taken from Menten \etal
 (\cite{menten}) is added.  The identification of the sources is presented
 on Fig.~\ref{Kp_dec_IRS}.

\begin{table}[!ht]
\caption {Offsets from \SgrA$^a$ of the sources shown on
Fig.~\ref{Kp_dec_IRS}.}
\begin{center}
        \begin{tabular}{|l|r| c c|}
\hline
ID & ID'$^b$ &$\Delta\alpha$cos$\delta$(\arcsec) & $\Delta\delta$(\arcsec)\\
\hline\hline
W &  40 & -4.17 & -1.96\\
E1 &  25 & -3.02 & -1.58\\
E2 &  27 & -3.24 & -1.68\\
E4 &  77 & -3.26 & -1.35\\
5 &  69 & -3.64 & -1.20\\
6 &  101 & -2.90 & -2.77\\
7 &  114 & -2.74 & -2.05\\
8 &  120 & -4.45 & -1.53\\
E3A &  118 & -3.24 & -1.48\\
10 &  145 & -4.35 & -1.19\\
E3B &  118 & -3.30 & -1.44\\
12 &  188 & -4.44 & -2.60\\
13 &  296 & -2.56 & -0.79\\
14 &  295 & -2.39 & -0.88\\
15 &  328 & -3.18 & -0.63\\
16 &  381 & -3.44 & -0.67\\
E5 &  780 & -3.46 & -1.50\\
18 &  902 & -2.55 & -1.42\\
E6 &      & -3.10 & -1.37\\
20 &  459 & -4.15 & -0.85\\
X  &      & -3.44 & -1.79\\

\hline

        \end{tabular}
\end{center}
\vspace{0.15cm}
\noindent
\\
a: \SgrA\/ position (Yusef-Zadeh \etal \cite{yusef}):\\
~~$\alpha$(2000) = 17~45~40.0383 $\pm$ 0.0007s \\
~~$\delta$(2000) = -~29~00~28.069 $\pm$ 0.014\arcsec \\
b: identification number from Ott \etal (\cite{ott2})\\

  \label{astrom}
\end{table}

   \begin{figure}[!ht]
\begin{center}
\resizebox{\hsize}{!}{\includegraphics{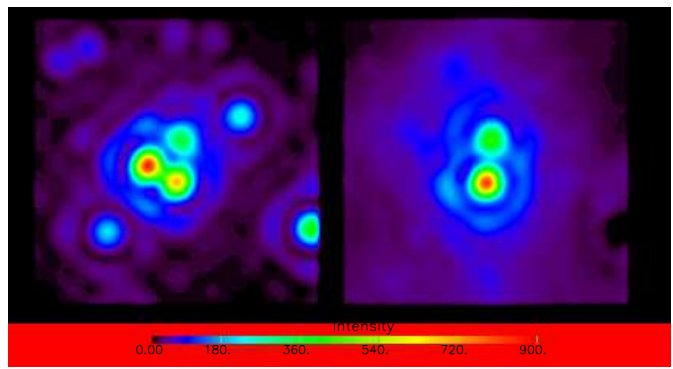}} \caption[]{At left,
      NICMOS images of IRS~13E in the Pa$\alpha$ filter (1.87\micron), and
      at right, difference between this image and the suitably scaled image
      in the continuum at 1.90\micron. Only IRS~13E2 and IRS~13E4 remain
      which appear as emission line stars.}  

\label{Paa}
\end{center}
   \end{figure}

\begin{table}[!ht]
\begin{center}
\caption{Projected separation of the main IRS~13E sources. From the
astrometry the precision on the distance betwen two sources is $\pm$ 15\,AU.}
        \begin{tabular}{| l  | c | c |} 
\hline
~~~Sources & Ang. sep. & Distance (AU) \\
\hline\hline
  ~~E1 $-$ E2 & 0.241\arcsec & 1930 \\
  ~~E1 $-$ E4 & 0.330\arcsec & 2630 \\
  ~~E2 $-$ E4 & 0.327\arcsec & 2600 \\ 
  ~~E2 $-$ E3A & 0.200\arcsec & 1600 \\
  E3A $-$ E3B & 0.073\arcsec & ~590 \\
\hline
        \end{tabular}
\end{center}

        \label{sepa}
\end{table}

 We used this astrometry to examine the position of the X-ray source CXOGC
 J174539.7-290029 whose IRS~13 is proposed as the optical counterpart by
 Baganoff \etal (\cite{baganoff}). The X-ray source was placed on
 Fig.~\ref{Kp_dec_IRS} by estimating the offset of the source with respect
 to \SgrA\/ from the coordinates of both sources reported in Baganoff
 \etal.  The 1-$\sigma$ error reported on the coordinates of each X-ray
 source is of $\pm$\,0.2\arcsec\ in right ascension and $\pm$\,0.1\arcsec\
 in declination. By combining the astrometric uncertainties on the two
 source positions we drew a minimum error box of $\pm$\,0.3\arcsec\ around
 the nominal position. The resulting position falls outside the center of
 IRS~13E, which will be discussed in Sect.~\ref{x-ray}.

\subsection{Photometry}

Table~\ref{detec} has been used to derive the standard photometry in H, K,
and L bands and the color indices of all the sources identified in
Fig.~\ref{Kp_dec_IRS}.  The Kp filter (Table~\ref{instrum}) which is close
and has a width comparable to K was used. It is referred to as K in the
results presented in Table~\ref{photom}.

\begin{table}[!ht]
\caption {H, K and L photometry of the IRS~13E cluster and of the nearby field
stars from Table~\ref{detec}.}
\begin{center}
        \begin{tabular}{|l|c c c |c|c|}
\hline
ID & $H$ & $K$ & $L$ & $H - K$ & $K - L$ \\
\hline\hline
W & 14.55 & 11.30 & 8.92 & 3.25 & 2.38\\
E1 & 12.74 & 10.98 & 8.59 & 1.76 & 2.39\\
E2 & 13.05 & 11.03 & 7.73 & 2.02 & 3.30\\
E4 & 14.37 & 11.72 & & 2.65 &\\
5 & 14.32 & 11.90 & & 2.42 &\\
6 & 14.71 & 12.13 & & 2.58 &\\
7 & 14.41 & 12.18 & & 2.23 &\\
8 & 14.86 & 12.23 & & 2.63 &\\
E3A & 16.48 & 12.47 & 7.50 & 4.01 & 4.97\\
10 & 15.55 & 12.80 & & 2.75 &\\
E3B & 16.95 & 13.07 & 7.92 & 3.90 & 5.14\\
12 & 15.92 & 13.28 & & 2.65 &\\
13 & 15.95 & 13.35 & & 2.60 &\\
14 & 15.95 & 13.42 & & 2.53 &\\
15 & 16.65 & 14.20 & & 2.45 &\\
16 & 16.66 & 14.22 & & 2.44 &\\
E5 & 18.19 & 14.28 & 8.48 & 3.90 & 5.79\\
18 & 16.79 & 14.52 & & 2.29 &\\
E6 & 16.24 & 14.56 & & 1.68 &\\
20 & 17.29 & 14.78 & & 2.52 &\\
\hline

        \end{tabular}
\end{center}
  \label{photom}
\end{table}

	 \subsection{Proper motions}
\label{propmt}

 From the results communicated by Ott \etal (\cite{ott2}), the direction
and amplitude of the proper motions for the main IRS~13E sources and most
of the stars in the 2.5\arcsec\/ field are represented in
Fig.~\ref{Kp_dec_IRS}. Note that the five sources, 13E1, 13E2, 13E3A, 13E3B
(proper motion is only given for the center of light of 13E3A and B) and
13E4 are all moving West, with a similar velocity, while all the nearby
stars have very different directions and amplitudes.  The amplitude of the
proper motions of the four main sources, in angular motion per year (for a
GC distance of 8\,kpc) and in velocity, are presented in
Table~\ref{mvprop}.  The reported uncertainties come from the mean error on
the proper motion vector coordinates adjusted on the set of positions. The
$rms$ uncertainty on 13E3 proper motion is the largest one, because of the
difficulty of measuring accurate positions from several epochs of SHARP
data for a weak source so close to much brighter sources, 13E2 and
13E4. The error becomes large for data recorded with poor seeing
conditions.

\begin{table}[!ht]
\caption {Amplitude of proper motions of the main IRS~13E
sources (from Ott \etal \cite{ott2}). }
\begin{center}
   \begin{tabular}{| c| c| l|}
\hline
  Name              &  mas/yr      &    ~~~km/s \\
\hline\hline
          13E1   &    5.50  &          207 $\pm$14 \\       
          13E2   &    8.20  &          310 $\pm$19 \\      
          13E3   &    7.54  &          285 $\pm$270 \\     
          13E4   &    7.77  &          294 $\pm$32 \\      
\hline
   \end{tabular}
\end{center}
\label{mvprop}
\end{table}

\subsection{The ionized gas near IRS~13E}
\label{iongas}
The \Paa\/ emission line is tracing also in the two central parsecs the HII
region called SgrA~West or commonly the \object{Minispiral}.  IRS~13E is
located just at the northern end of a bright emission arc, part of the
circular structure open toward \SgrA, called the Minicavity.  In Stolovy
\etal (\cite{stolovy}) the bubble-like feature is proposed as created by
the wind from a star identified by a weak Pa$\alpha$ point source, since
located very near the geometric center of the Minicavity. A proper motion
of the edge of the Minicavity of $\sim$\,200\kms\/ is measured by Zhao \&
Goss (\cite{zhao}). In Paumard \etal (\cite{paumard2}), it is shown that
the Minicavity is embedded inside a non-planar gas flow of the Minispiral,
called the \object{Northern Arm}. In the field of the Minicavity another
velocity structure is identified, stretched toward northwest, called the
\object{Bar}, roughly perpendicular to the bright edge of the Northern Arm.
On the line of sight of IRS~13E these two velocity structures produce the
two main components in the line profile of the ionized gas shown on
Fig.~\ref{HeI_Brg} central panel. The fastest component at -250\kms\/ is due
to the motion of the gas disturbed by the Minicavity, and the slowest
component, at -39\kms, to the Bar. On a morphological basis IRS~13E would
seem to belong to the bright arc of the gas shocked at the edge of the
Minicavity.  On the intensity map of the Bar, isolated by this
multi-component analysis, a small region just centered on the IRS~13E
position appears locally enhanced.  Therefore, the observed brightness of
the ionized gas around the position of IRS~13E is due to the addition of
two contributions on the line of sight: the edge of the Minicavity and the
locally excited gas of the Bar. Only this latter component is due to the
presence of the IRS~13E sources, by their strong ionizing flux.
Hence, IRS13E should be located close to or inside the Bar, which lays
behind the Minicavity (Paumard \etal \cite{paumard2}).

\subsection{The stellar \Paa\/ emission line}

As shown on Fig.~\ref{Paa}, only two of the IRS~13E components,
IRS~13E2 and 13E4, remain after subtraction of the F190N continuum
from the F187N filter which contains \Paa.  These two stars are
unambiguously emission line objects, with the integrated line intensity
at IRS~13E2 brighter by a factor 2.35 $\pm$\,0.1 than at
\object{IRS~13E4}.

\subsection{The Br$\gamma$ and the \ion{He}{i} 2.058\micron\/ line profiles}

\begin{figure}[!ht]
\begin{center}
\resizebox{\hsize}{!}{\includegraphics{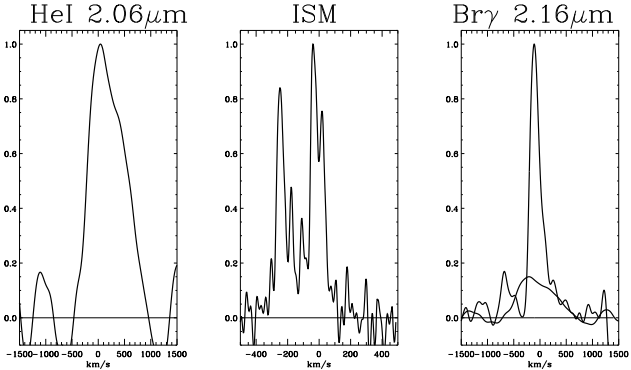}}
\caption[]{\ion{He}{i} 2.06$\mu$m and Br$\gamma$ emission line profiles at
IRS~13E normalized to 1, from the BEAR data, after subtraction of the
interstellar emission (ISM) contribution the observed profiles. The
difference of linewidth is noticeable. The multi-component ISM emission,
extracted from a ring around IRS~13E in the Br$\gamma$ data cube, is shown
in the middle panel. Since \ion{He}{i} 2.06$\mu$m is coming only from 13E2
(Cl\'enet \etal \cite{clenet2}) the broad profile adjusted on the wings of
the Br$\gamma$ profile, of same width as the \ion{He}{i} profile, shows the
relative intensity of Br$\gamma$ from 13E2 compared to 13E4.}
\label{HeI_Brg}
\end{center}
   \end{figure}

The profiles of these two emission lines were obtained from spectra
extracted on the same aperture size (3$\times$3 pixels i.e. $\simeq
1\arcsec\times1\arcsec$) from each BEAR data cube, at the IRS~13E
position. With the spatial resolution of the BEAR data the contribution of
the two emission line stars 13E2 and 13E4 is mixed, to which is added the
ISM emission.  After the most plausible subtraction to each line profile of
this latter contribution, shown on the central panel of Fig.~\ref{HeI_Brg},
possible thanks to the spectral resolution of the data, Br$\gamma$ appears
much narrower than the \ion{He}{i} 2.06\micron\/ line. With a FWHM of
215\kms, compared to $\sim$\,900\kms\/ for the helium line, as already
measured in Paumard \etal (\cite{paumard}), the two lines should belong to
two different sources, which is discussed in Sect.~\ref{e2_e4}.

\subsection{Spectral energy distribution and extinction law}

\begin{figure}[!ht]
\begin{center}
\resizebox{\hsize}{!}{\includegraphics{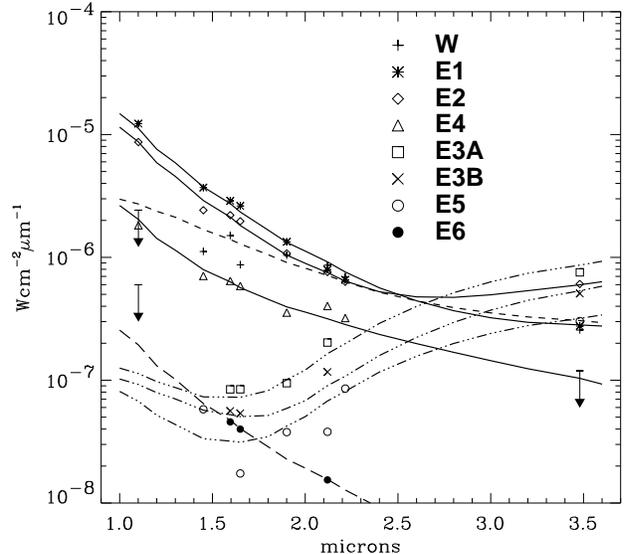}}

      \caption[]{Dereddened SED of the IRS~13 sources in
      W\,cm$^{-2}$\,\micron$^{-1}$.  The top of the arrows indicates the
      estimated upper limit of the detectable flux in the 1.1\micron\/ and the L-band
      filters.  The various lines represent the best fit between 1 and
      4\micron\/ of the data points from a two-component model with the
      parameters of Table~\ref{dered} and the most adapted $A_v$ values.}

         \label{sp13E_dered}
\end{center}
   \end{figure}

The photometric measurements presented in Table~\ref{detec} were used to
build the SED between 1 and 4\micron\/ of the IRS~13 components
(Fig.~\ref{sp13E_dered}), and also of all the stars detected in the
surrounding field.  To achieve the final goal of determining the spectral
type of these stars, a dereddening has to be applied over this range.  The
extinction over the central parsecs is highly variable (Blum \etal
\cite{blum96}, Rieke \cite{riekem}, Scoville \etal \cite{scoville}). We
took the most recently published law, which is a merging of previous works
(Moneti \etal \cite{moneti}), and adjusted $A_v$, making first the
simplifying assumption that the reddening toward IRS~13E and the few
arcsecs around would be identical for all the sources. For this adjustment
we were helped by two constraints: IRS~13E2 and 13E4 are emission line
sources (Fig.~\ref{Paa}), therefore hot sources with a blue SED, while
IRS~13W is a cool star (Krabbe \etal \cite{krabbe}).  Black-body curves
were fitted to the data. $A_v$ was adjusted in order to fulfill the two
constraints. Higher values of $A_v$ make the sources bluer, while lower
values make the sources redder. Temperatures higher than 25,000\,K had to
be introduced for IRS~13E1, 13E2 and 13E4, which correspond to the
Rayleigh-Jeans regime in this spectral range. In that case, only a lower
limit of T$_{eff}$ can be derived. The slope of the SED becomes constant in
a log[F($\lambda$)] diagram. For these stars the adjustment of $A_v$ makes
it possible to bring the SED parallel to the data points, providing the
strongest constraint on $A_v$. Finally, a value of $A_v = 35~\pm\,0.5$ was
adopted from the three hot stars, 13E1, 13E2 and 13E4. A mean value of $A_v
=~30.5$ had been determined by Rieke (\cite{riekem}) from a survey of the
stars in the central parsec, which excluded objects like IRS~13 from the
color-magnitude diagram because of the surrounding dust. We confirm a much
higher $A_v$ value for IRS~13E. On the other hand, for IRS~13E6 $A_v = 35$
was too high for a good fit to the data. A value of 29 was more
appropriate. On the contrary, we will show in Sect.~\ref{irs13w} that a
value of $A_v = 38.5$ had to be adopted for IRS~13W.

However, it appeared that the fit of the dereddened data from
Table~\ref{detec} was not possible with a single temperature for the IRS~13
sources. Except 13E6, they show an infrared excess. Then, the fit was made
as a sum of two black-body curves, $Coef_1\times BB(T_1) + Coef_2\times
BB(T_2)$. The four parameters of adjustment obtained for each star are
reported in Table~\ref{dered}. The final SEDs are shown on
Fig.~\ref{sp13E_dered}. From all the results presented above a spectral
classification of the stars detected in the IRS~13E cluster is proposed.

\begin{table}[!ht]
\caption {Fitting parameters of the SED of the IRS~13 sources.  The
spectral type of each source, as discussed in Sect.~\ref{nature}, is
summarized in the last column.}
\begin{center}
\begin{tabular}{| l@{} | l@{}  r  r  r l | c |}
\hline
Star & {\sl Coef$_1$} & {\bf T$_1$}~K & {\sl Coef$_2$} &
 {\bf T$_2$}~K & ~A$_v$ &{\bf Sp. Type} \\
\hline

W  &{\sl  13.50} &{\bf  3600} &{\sl  4250} &{\bf  640}&          38.5&M3III  \\
E1 &{\sl  0.800} &{\bf $\geq$ 25000} &{\sl  12000} &{\bf  550}& 35&$\sim$ O5I \\
E2 &{\sl  0.620} &{\bf $\geq$ 25000} &{\sl  40000} &{\bf  550}& 35& $\sim$ WN8\\
E4 &{\sl  0.140} &{\bf  $\geq$ 25000} &{\sl  45} &{\bf 1550}&35&$\sim$ O5IIIe \\
E3A &{\sl 0.460} &{\bf  3800} &{\sl  33000} &{\bf  610}& 35& d. WR$^a$ \\
E3B &{\sl 0.375} &{\bf  3800} &{\sl  29000} &{\bf  580}& 35& d. WR$^a$ \\
E5 &{\sl  0.070} &{\bf  6000} &{\sl   9800} &{\bf  630}  & 35& d. WR$^a$ \\
E6 &{\sl  0.013} &{\bf $\geq$ 25000} & &                & 29& $\sim$ O5V   \\
\hline
   \end{tabular}
\end{center}
\vspace{0.15cm}
\noindent
\\ 
a~: dusty Wolf-Rayet star\\

\label{dered}
\end{table}

\section{Nature of the IRS~13 sources}
\label{nature}
        \subsection{IRS~13E1}

This source is characterized by:
\begin{itemize}
\item [--] no emission detected in Pa$\alpha$ (Fig.~\ref{Paa}),
\item [--] an elongated envelope in H and Kp toward 13E2, which
shows at the star position a FWHM of
180\,mas (to compare to the 40\,mas of the PSF) or 1435~AU,
\item [--] a SED fitted by a $\geq$\,25,000\,K source and an infrared excess at
550\,K (Table~\ref{dered}),
\item [--] a K$_{mag}$ of 10.9 (Table~\ref{photom}).  At the distance of the GC
and with the adopted extinction, IRS~13E1 is too luminous to be a dwarf
star. A O5V star of comparable $T_{eff}$ would have a K$_{mag}$\,$\simeq$\,14
(with A$_K$\,$\simeq$\,4). IRS~13E1 is $\sim$\,3 magnitudes brighter.
\end{itemize}

Hence, IRS~13E1 must be a blue, supergiant star. The surrounding halo of
scattered emission is indicative of a strong stellar wind, which also
favors the identification of a massive, hot star.  Consequently, IRS~13E1
might be close to a O5I spectral type.  However, at the position of the
source there is no evidence in the Pa$\alpha$ image of a negative level
(Fig.~\ref{Paa}), which would be the signature of the photospheric hydrogen
line in absorption since no emission is detected. Possibly, with the
uncertainty on the adjustment of the continuum subtraction from the
1.87\micron\/ filter, the integrated emission from the envelope in the
image can be compensating the photospheric absorption. Only a near-infrared
spectrum of IRS~13E1 behind adaptive optics could confirm this assumption.

        \subsection{IRS~13E2 and 13E4}
\label{e2_e4}
 IRS~13E2 and 13E4 are two emission line stars (Fig.~\ref{Paa}).
 In the K band IRS~13E2 is brighter than IRS~13E4 by a
 factor $\simeq$\,2 (Table~\ref{photom}).  In our L band 13E4 is no longer
 detectable while 13E2 can still be measured.  With this detection it turns
 out that all the emission spectra of IRS~13E obtained up to now (see
 references in Introduction) are in fact the combination of two emission
 line stars which are likely different.

A decisive element is reported by Cl\'enet \etal (\cite{clenet2}). From
spectro-imaging with a FP in the 2.06\micron\/ helium line behind the
CFHT-AO system they indicate that IRS~13E2 is the only helium emitter.  The
radical difference of linewidth between the \ion{He}{i} 2.06\micron\/ and
Br$\gamma$ stellar line profiles at IRS~13E (Fig.~\ref{HeI_Brg}) brings
another piece of information.  A similar difference of linewidth of the
\ion{He}{i} 2.06\micron\/ lines is observed among the hot stars of the
central cluster (Paumard \etal \cite{paumard}, \cite{paumard3}), leading to the
identification of two classes of massive stars. Applying this criterion to
13E2 and 13E4, and taking into account the detection of \ion{He}{i}
2.06\micron\/ only at IRS~13E2, it can be concluded that IRS~13E2 as a
strong helium emitter with broad line is a late-type WR star, probably of
WN type (Figer \etal \cite{figer}) since no detection of \ion{C}{iii} or
\ion{C}{iv} typical of WC type is reported from earlier K-band spectra. A
$T_{eff}$ $\geq$\,25,000~K estimated for the source (Table~\ref{dered}) is
consistent with this identification. IRS~13E4 with also a $T_{eff}$
$\geq$\,25,000 K (Table~\ref{dered}), but no helium emission, source of the
narrower Br$\gamma$ profile, is more likely a less evolved star than
IRS~13E2.  From its SED and its absolute brightness in K, fainter than
IRS~13E1, IRS~13E4 can be reasonably proposed as a O5IIIe star or just
reaching the LBV stage but in a high extinction phase to explain its weakness.

However, there is an apparent contradiction between the brightness of
IRS~13E2 in \Paa\/ in Fig.~\ref{Paa} and the Br$\gamma$ profile at IRS~13E
from the BEAR data after correction for the ISM emission
(Fig.~\ref{HeI_Brg}).  As \Paa\/ appears strong at IRS~13E2 in
Fig.~\ref{Paa}, the Br$\gamma$ profile at IRS~13E should be dominated by
the emission from this source and be as wide as the \ion{He}{i}
2.06\micron\/ profile. That is not what is observed as shown on
Fig.~\ref{HeI_Brg}. To reconcile these two facts, it must be noticed that
\Paa\/ is far from a perfect indicator with narrow-band imaging technique,
to distinguish between hydrogen-rich and helium-rich emitters since the
\Paa\/ line (1.8751\micron) is blended with a strong helium line,
\ion{He}{i}~(4-3) at 1.8697\micron. There is another helium line within the
bandpass of the continuum filter, at 1.9089\micron, but which is weak and
will contribute to subtract only a little of the helium emission. All these
features are well seen in the CGS4 spectrum of the AF star around \Paa,
presented by Najarro \etal (\cite{najarrob}), which is another helium star
belonging to the class of the broad-line stars (Paumard \etal
\cite{paumard}, \cite{paumard3}).  Hence, the intensity in the F187N-F190N
image at the star position cannot be considered as a fully reliable
measurement of the true \Paa\/ emission in the stellar atmosphere. The
bright spot at IRS~13E2 in Fig.~\ref{Paa} is likely due to the \ion{He}{i}
1.8697\micron\/ line and with some contribution of the \Paa\/
emission. That is consistent with the Br$\gamma$ profile shown on
Fig.~\ref{HeI_Brg} which can be decomposed into a narrow line, whose origin
must be IRS~13E4, and a fainter, broad component of same width as the
\ion{He}{i} 2.06\micron\/ line, which should be the contribution of
IRS~13E2 to the observed profile.  The residual Br$\gamma$ profile will
show a P~Cyg profile, typical of hot stars with an atmosphere in expansion,
consistent with the spectral type attributed to IRS~E4.

\subsection{IRS~13E3A and B}

IRS~13E3 resolves into a double source in the deconvolved AO images in H
and Kp (Fig.~\ref{Kp_dec_IRS}). Their projected separation is equal to
$\simeq$\,600~AU (Table~\ref{sepa}).  The two sources are photometrically
quite identical. They are extremely red objects as indicated by the
measurements in Table~\ref{photom}, and from their SED
(Fig.~\ref{sp13E_dered}). They are faint in the H band and prominent in the
L band. We measure a K-L color index of $\sim$\,5 mag. for the two
components (Table~\ref{photom}). Several sources in the inner parsec,
mainly located along the Northern Arm (\object{IRS~1W}, 2, 3, 5, 10W, 21),
are also very red objects, with K-L $>$ 3, reported in Cl\'enet \etal
(\cite{clenet}).  IRS~1W and \object{IRS~21} have featureless spectra in
the K band (Blum \etal \cite{blum96}). IRS~21 has been studied in detail,
from 2 to 25\micron\/ by Tanner \etal (\cite{tanner}).  They have fitted
its SED by a two-component model, the near-infrared scattered light from
the central source peaking at $\simeq$\,3.8\micron\/ (760~K), and the
mid-infrared re-emitted light from the dust shell at $\sim$\,250~K. They
conclude that IRS~21 is a dusty WR star, experiencing rapid mass loss as
well as the other luminous Northern Arm sources (Tanner \etal
\cite{tanner2}). The IRS~13E3 SEDs are also fitted by two components
(Table~\ref{dered}), but with respectively 3800~K and
600~K. \object{IRS~13E3A} and E3B are likely sources of the same type as
IRS~21 and the other Northern Arm sources. The higher temperature of the
infrared component can come from the additional heating of the dust shell
by the very close, massive blue stars, IRS~13E1, 13E2 and 13E4. That is
also consistent with IRS~13 being not a prominent source at 12.5\micron\/
on the images in this band (Tanner \etal \cite{tanner}) compared to the
other Northern Arm sources.

  \subsection{\object{IRS~13E5}}

 One of the sources revealed by deconvolution of the AO images
(Fig.~\ref{Kp_dec_IRS}) we propose to name IRS~13E5, is also present in
the SHARP data (Ott \etal \cite{ott2}). From the dereddened photometry
(Fig.~\ref{sp13E_dered}) this source has a SED similar to IRS~13E3A and
E3B, being roughly a factor 2 fainter than each of the IRS~13E3
components. Its SED is also fitted by two thermal components, with
temperatures of the same order as for 13E3A and E3B
(Table~\ref{dered}). From this similarity we propose that IRS~13E5 is
another example of dusty WR star, possibly more embedded, behind IRS~13E3A
and E3B.

    \subsection{\object{IRS~13E6}} 

IRS~13E6 is detected in H, in the F160W filter, a broader H filter, and
near the detection limit of the Kp band (Table~\ref{detec}) and not in L.
It is not detected by Ott \etal (\cite{ott2}) whose data come from K-band
imaging on the NTT.  With $T_{eff} \geq$\,25,000~K and K$_{mag}$ = 14.56
(Table~\ref{photom}) 13E6 is a weak, hot star, 3.5 magnitudes fainter than
IRS~13E1. It can be considered close to a main sequence O5V star. On
Fig.~\ref{sp13E_dered} the best fit to the data was obtained with a value
of $A_v$ lower than for the other IRS~13E sources. It should mean that
IRS~13E6 is located on the line of sight, but in front of the IRS~13E
complex.

     \subsection{The red halo around IRS~13E} 
\label{dif_hal}
\begin{figure}[!ht]
\begin{center}
\resizebox{\hsize}{!}{\includegraphics[bb=70 158 446 521]{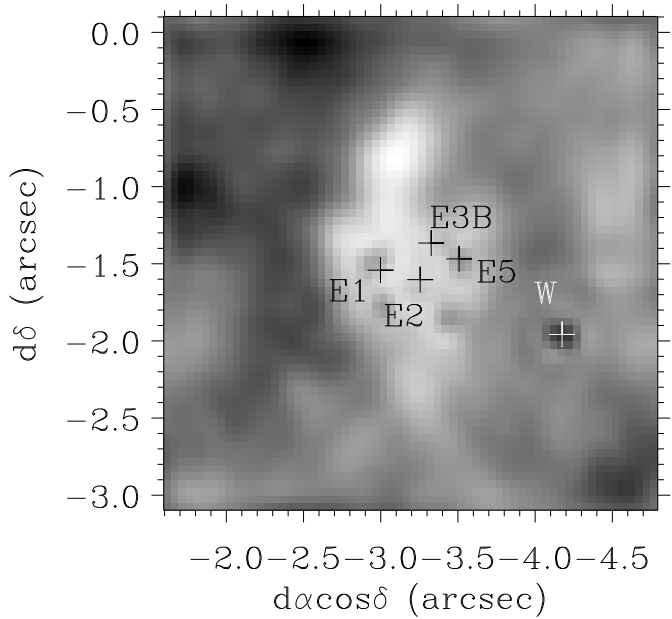}}
   \caption[]{Map of the residual fitted continuum in the deconvolution
   operation of the L-band image with the MCS code.  The identified stars
   in this band (Table~\ref{detec}) have been subtracted explaining the
   dark holes in the map (e.g. at the W position). Source positions are
   marked by crosses with their names. To limit confusion, for the E3
   binary only E3B position is indicated. The apparent halo might be not
   purely thermal emission of dust and contain more fainter embedded
   sources (Eckart \etal \cite{eckart2}). An estimation of the halo
   contrast is given for the brightest pixel (position $-$~3.1\arcsec,
   $-$~0.8\arcsec), which is four times brighter than east of 13E, out of
   the halo.}
 
   \label{fd}
\end{center}
   \end{figure}

The flux in the L band is only due to continuum emission since the
Br$\alpha$ line from the ionized gas falls outside the filter bandpass.
The deconvolution of the AO image in this band, after subtraction of the
stars, leaves an enhanced background emission around IRS~13E shown on
Fig.~\ref{fd}, extending on 2\arcsec in the North-South direction, above an
almost uniform continuum. On a larger scale, over the central parsecs, a
diffuse emission is seen in the mid-infrared (Tanner \etal \cite{tanner})
whose structure is exactly following the intensity map of the ionized gas
forming the Minispiral (Stolovy \etal \cite{stolovy}, Morris \& Maillard
\cite{morris}, Paumard \etal \cite{paumard2}). This emission is the thermal
emission from the dust dragged by the same fast flowing motion as the gas,
heated through the trapping of Ly$\alpha$ photons from the central, massive
star cluster (Rieke \etal \cite{rieke}).  In the analysis of the local
structure of the interstellar gas (Sect.~\ref{iongas}) we have shown that
the IRS~13 complex is embedded in the flow called the Bar (Paumard \etal
\cite{paumard2}). This flow is also likely mixed with dust.  Hence, the
enhanced continuum emission seen in L around IRS~13E should come from the
dust heated up by the strong UV field coming from the local concentration
of hot stars. There, the emission temperature of the dust has a value
around 600\,K from Table~\ref{dered}, against 250\,K at IRS~21 (Tanner
\etal \cite{tanner}) and probably lower out of the embedded sources.  From
further observations with NAOS/CONICA in the same band (Eckart \etal
\cite{eckart2}) the bright extension 1\arcsec\/ north of IRS~13E resolves
into a compact cluster of several very red sources. Their spectral type
might be comparable to the dusty WRs as 13E3A, B and 13E5 detected next to
it. Spectroscopic studies are needed to confirm this hypothesis.

 \subsection{IRS~13W}
\label{irs13w}
 This star is associated with IRS~13 only for historical reasons.  We kept
it in all the study since its cool stellar type, confirmed by the presence
of CO in its K-band spectrum (Krabbe \etal \cite{krabbe}), was a constraint
on the determination of the local value of $A_v$.  In the deconvolved
Kp-band image 13W shows a larger size than the PSF which might be
indicative of a dusty envelope, attested also by a significant infrared
excess (Table~\ref{dered}). However, in the first analysis with the
assumption of the same $A_v$ value for all the IRS~13 sources a $T_{eff}$
of $\sim$~2600~K was obtained for 13W. With this temperature the cool star
should be of Mira-type. A H+K spectrum would show the deep absorption of
water vapor on each side of the K band. With the NICMOS data we have a
photometric measurement at 1.90\micron, just in the water vapor
band. Fig.~\ref{sp13E_dered} shows a smooth distribution of data
points. Thus, we must conclude that this temperature is too low. To
increase this temperature in the fit a higher $A_v$ value must be assumed
locally. With the four parameters of the fit (Table~\ref{dered}) plus $A_v$
the solution is not unique. A plausible solution is obtained with
$A_v$~=~38.5 and $T_{eff}$~=~3600~K. A higher temperature would require a
higher $A_v$ value which would become inconsistent with the non detection
of IRS~13W at 1.1\micron\/ (Table~\ref{detec}). With this $T_{eff}$
temperature and its luminosity IRS~13W can be assumed to be a M3 giant
star.

  \subsection{The other stars of the IRS~13 field}
\label{ostar} 
As presented in Table~\ref{detec} twelve more stars are detected in the
small field around IRS~13.  Note that they are only detected in the
1~-~2.3\micron\/ range. They offer an opportunity to study in detail a
small sample of the central parsec stellar population. By fitting
black-body curves (a single temperature component was required), under the
assumption of the same $A_v$ as for IRS~13E, nine stars are main sequence K
and M stars, with $T_{eff}$ from 2800 to 5000\,K. Hence, they belong to the
numerous old star population of the GC (Blum \etal \cite{blum96}). The
three remaining stars ($\#$\,7, 12, 18 - Fig.~\ref{Kp_dec_IRS}) are very
hot stars, with $T_{eff}$ $\geq$\,25,000~K and K$_{mag}$ respectively of
12.2, 13.3 and 14.5 (Table~\ref{photom}).  As proposed for IRS~13E6, they
should be main sequence O stars. Near infrared spectroscopy is needed to
better constrain their spectral type. However, in this small field four out
of a total of 14 sources (putting the massive components of IRS~13E apart)
are apparently O-type stars, which could make a proportion of about one
quarter of such stars in the central parsec, the rest being late-type stars.

\section{IRS~13E as the remaining core of a massive star cluster}

 IRS~13E appears as only composed of hot, massive stars.  This
 concentration within 0.5\arcsec\/ cannot be fortuitous.  Further deep
 high-spatial resolution imaging could eventually reveal more components
 (Eckart \etal \cite{eckart2}).  The common westward direction and similar
 amplitude of the proper motions with a mean value of $\sim$\,280\kms\/ for
 the main components (Sect.~\ref{propmt} and Table~\ref{mvprop}) is a
 decisive argument to indicate that 13E1, 13E2, 13E3A/B and 13E4 are
 physically bound. Regarding 13E5 for which a proper motion cannot be
 currently measured to associate it unambiguously to the cluster, its
 spectral type, identical to 13E3A/B, leads us to conclude that this source
 belongs to IRS~13E too. For IRS~13E6 the proper motion is not available
 either, but the value of $A_v$ 6 magnitudes lower than for IRS~13E likely
 indicates that the source does not belong to the cluster. The source is
 another O-type star, just located on the line of sight.  However, in
 conclusion, the source historically called IRS~13E is a compact, massive
 star cluster with at least six members. That also means a young star
 cluster of a few 10$^6$~yr old, since several members are identified as
 having already reached the WR stage.

 The compactness of the cluster and the common proper motion of the
 components raise the question of the force which keeps the massive stars
 bound. The hypothesis of the presence of a dark, massive source, a stellar
 black hole at the center of IRS~13E is natural. An intermediate-mass black
 hole (IMBH 10$^3$ to 10$^4$~M$_{\odot}$) is supposed to form by runaway
 growth in massive, young stellar clusters as a result of stellar
 collisions in the cluster center, as was modeled by Portegies Zwart \&
 McMillan (\cite{zwart2}). Constraints on the possible central mass could
 be obtained from the radial velocities of the sources. It is estimated
 only for the two emission line stars, with a positive velocity of
 $\simeq$\,30\kms\/ for IRS~13E2 and a negative velocity of
 $\simeq$\,30\kms\/ for IRS~13E4 (Fig.~\ref{HeI_Brg}).  Associated with
 parallel and equal proper motion vectors (Fig.~\ref{Kp_dec_IRS}), this
 suggests that both stars orbit around the center of mass in a plane
 orthogonal to the plane of the sky.  Assuming that both stars orbit around
 the black hole in a symmetrical fashion on circular orbits, half the
 projected separation between IRS~13E2 and IRS~13E4 (Table~\ref{sepa})
 gives an orbit radius of 1300~AU$/\cos{i}$ where $i$ is the angle that the
 line containing both stars makes with the plane of the sky. The radial
 velocity of each star ($\simeq30$~km~s$^{-1}$) can be used as an estimate
 for their orbital velocity. A period of rotation of $\simeq$\,1295\,yr is
 derived. Then, the 3rd Kepler's law gives directly a total mass of
 $\simeq1300$~M$_{\odot}\cos^{-1}(i)R_0/8$~kpc. We can also release the
 circular orbits hypothesis, and only assume that the system is bound,
 which means that the potential energy of each stars is greater than its
 kinetic energy. This gives a lower limit to the black hole mass, half the
 previous estimate: $M_{BH}>750$~M$_\odot\cos^{-1}(i)R_0/8$~kpc. The
 dependence of these two values is quadratic on the orbital velocity of the
 stars and linear on their distance to the black hole, so that the
 constraint $M_{BH}\geq~10^3$~M$_{\odot}$ can be considered rather robust.
 It falls within the range derived by Portegies Zwart \& McMillan
 (\cite{zwart2}) for a black hole formed in the core of a dense star
 cluster with massive stars of initial mass $\geq$\,50\,M$_{\odot}$.

 	\subsection{The fate of a star cluster near \SgrA}

The large number of massive stars in the central parsec, which are very
rare elsewhere in the Galaxy, remains one of the major mysteries of this
region. Since star formation would be difficult due to the strong tidal
forces from the \SgrA\/ black hole, Gerhard (\cite{gerhard}) made the
interesting hypothesis that the central parsec \ion{He}{i} stars, the most
prominent of the massive young stars, might be the remains of a dissolved,
young cluster, which originally formed further away from \SgrA. He argued
that the Arches and the Quintuplet clusters, located within a projected
distance of $\sim$~30pc from \SgrA\/, testify that star formation by
cluster of massive stars has been occurring in the nuclear disk of the
Galaxy.  If one examines Table~\ref{dered}, IRS~13E appears as a kind of
summary of all the spectral types of young stars observed in the central
parsec from O to WR. Among the helium stars only IRS~13E appeared multiple,
which motivated the current study. With the hypothesis that IRS~13E might
be the remaining core of a massive star cluster, was this cluster the source
of the population of massive stars observed in the central parsec?

Morris (\cite{morris93}) has argued that it would take longer than the
lifetime of massive young stars to transport them inward within the central
parsec if they formed at too large distance. The same argument is applied
again by Figer \etal (\cite{figer00}), who claim that the \SgrA\/ cluster
(Genzel \etal \cite{genzel97}, Ghez \etal \cite{ghez}, Gezari \etal
\cite{gezari}) could not have formed more than 0.1~pc from the center, and
then, the initial clump should have an exceptional density
$\geq$\,4\,10$^{11}$\,cm$^3$.  Gerhard (\cite{gerhard}) discarded this
argument by a revision of the conditions for a cluster formed at 30~pc to
spiral into the center within the lifetime of its most massive stars. The
main condition is that the initial mass of the parent cloud in which the
cluster forms must be massive enough ($\simeq$\,2$\times$10$^6$M$_{\odot}$)
to survive the evaporation in the strong tidal field of the nuclear
bulge. He concludes that clusters significantly more massive than the
Arches cluster and formed a little closer than 30~pc can reach the central
parsec in due time. In order to test this statement Kim \& Morris
(\cite{kim}) have made several simulations, for different masses
(10$^5$M$_{\odot}$ and 10$^6$M$_{\odot}$) and different initial orbit radii
(2.5 to 30~pc), of the dynamical friction on a star cluster near
\SgrA. They came to the conclusion that some simulations can be regarded as
candidates for the origin of the central parsec cluster, but that \lq\lq
the required conditions are extreme\rq\rq\/, with an initial mass of the
cluster of 10$^6$M$_{\odot}$ or a very dense core
$\geq$~10$^8$M$_{\odot}$pc$^{-3}$ (Kim \etal \cite{kim2}). A mass of the
cluster of 10$^6$M$_{\odot}$ supposes a very large number of particles,
with some of initial mass $\geq$~10$^2$M$_{\odot}$ to reach rapidly the WR
stage. Compared to the relatively small number of detected helium stars,
concentrated in the central parsec, $\sim$~19 from the revision by Paumard
\etal (\cite{paumard3}), plus few more dusty WRs, a very large quantity of
O-type stars ($>$~10$^5$) should be detected, which is not the case from
the proportion of such stars we count in the IRS~13 test field (Sect.~\ref{ostar}).  This might seem an argument
against this scenario.

However, another analytic work by McMillan \& Portegies Zwart
 (\cite{zwart}) has reconsidered the fate of a star cluster near the
 central dark mass. They tried to address the problem more completely by
 taking into account, in addition to the initial mass and the distance to
 the center, the original mass function of the cluster, the initial cluster
 radius and the stellar evolution through mass loss during the inspiral
 time of the cluster. They conclude that star clusters born with masses
 $\simeq$\,10$^5$M$_{\odot}$ within 20~pc from the center, with a half-mass
 radii of $\sim$\,0.2~pc can reach a final distance of 1~pc within 10\,Myr.
 As a secondary conclusion, they assess that from their mass and their
 distances, the Arches and the Quintuplet clusters, will never reach the
 vicinity of \SgrA. This latter work makes the origin of the central,
 massive star cluster by the dissolution of a compact cluster in the
 galactic tidal field more plausible, not requiring extreme mass conditions
 as in the simulation of Kim \etal (\cite{kim2}).  In conclusion, we
 propose that the IRS~13E cluster, by its unique location and composition,
 is the possible core of an earlier massive star cluster, formed about
 10\,Myr ago, within 20\,pc of \SgrA, with a mass of
 $\simeq$\,10$^5$M$_{\odot}$, which was the progenitor of the entire hot
 star population, from WR to O-type stars, observed today in the central
 parsecs of the Galaxy. In addition, the hypothesis of IRS~13E harboring a
 IMBH as consequences on this scenario.

	\subsection{The need for a second black hole}

A recent paper by Hansen \& Milosavljevi\'c (\cite{hansen}) came to
our attention when the current work on IRS~13 was completed. The authors
try to solve the difficulty of bringing disparate groups of hot stars, the
WR-type stars, the LBV-type stars and the S-cluster, at their observed
location within the timescale required within a single evolutionary
scenario. They argue that massive star clusters can sink within the
required star lifetime but are tidally disrupted at a distance greater than
1 parsec from \SgrA, from which it would result a population of sources
with low binding energy orbits unlike those of the helium stars and
particularly of the S-cluster stars orbiting within 0.1\,pc of \SgrA*. For
this purpose they propose a model with an infalling IMBH. The stellar
orbits continue to evolve by undergoing close encounters with the IMBH, bringing
some stars near enough to be trapped by the massive BH. This model appears as a
refinement of the same idea of an origin of the young star population in
the central parsec as formed at a distance of the \SgrA\, $>$ 10\,pc where
star formation can occur, in a dense star cluster sinking towards \SgrA. We
propose IRS~13E with a IMBH as the possible remnant of this initial cluster.

\section{What is the source of the X-ray emission at IRS~13?}
\label{x-ray}
The origin of the X-ray emission at IRS~13 reported by Baganoff \etal
(\cite{baganoff}) for the position, by Coker \etal (\cite{coker}) for the
spectrum, and by Muno \etal (\cite{muno}) for complementary data on the
energy distribution, must be examined in the context of the nature of
IRS~13E presented in the current work.  According to Baganoff \etal the
X-ray source is located 0.56\arcsec\/ northeast of IRS~13W. Using our own
astrometry (Table~\ref{astrom}) and the offset of the IRS~13 X-ray source
from the X-ray source at \SgrA, the source falls 0.75\arcsec\/ east of
IRS~13W (Fig.~\ref{Kp_dec_IRS}). With some minor difference we agree that
the nominal position of the X-ray source does not coincide with the center
of IRS~13E (our position is 0.37\arcsec\/ south).  This discrepancy is
larger than the error box on the X-ray source. Then, the source is
apparently not coincident with either IRS~13E2 as a post-LBV binary as
originally proposed by Coker \etal (\cite{coker}), or the IRS~13E cluster
itself, which would have been the most plausible candidate with its rare
concentration of several massive, hot stars with high mass loss and very
fast winds, and possibly with a black hole at its center.  Except a better
astrometry the source CXOGC J174539.7-290029 must be considered as
a source independent of IRS~13E. A systematic X-ray survey of the
Galactic Center region at sub-arcsecond scale with Chandra by Muno \etal
(\cite{muno}) over a field of 17\arcmin$\times$17\arcmin\/ centered on
\SgrA\/ has revealed more than 2000 discrete X-ray sources.  Stellar
remnants, white dwarfs with magnetically accreting disks, binaries with
neutron stars or solar-mass black holes are considered as responsible for a
large fraction of these discrete sources. Such a source, too weak in the
near infrared to be detected, could be present on the line of sight of
IRS~13, since as measured by Muno \etal (\cite{muno}) the distribution of
the density of discrete sources peaks in the central parsec. The binary
explanation for many stellar X-ray sources is based on a distinct emission
feature centered at $\sim$\,6.7\,keV and otherwise a featureless
spectrum. The spectrum presented by Coker \etal (\cite{coker}) stops at
6\,keV, probably because no energy was detected beyond, but it shows a
narrow emission line at 3\,keV.  Wang \etal (\cite{wang}) indicate that
this line should be due to the contribution of massive stars. Hence, the
infrared counterpart to the X-ray source could be detected by deeper
spectrophotometry of the IRS~13 field.  A good S/N ratio X-ray spectrum of
the so-called IRS~13 source would also help to better constrain the nature of
this source.

\section{Conclusion}

The presence of a compact star cluster of six hot, massive stars at the
position of IRS~13E from high-resolution near-infrared observations is
demonstrated. The spectral types of the various members range from O to WR,
including dusty WRs. Proper motion measurements indicate that the brightest
stars are co-moving suggesting that the members of the cluster are bound by
a central IMBH with a mass $\geq$\,1300\,M$_{\odot}$.  Such a secondary
black-hole in the vicinity of \SgrA\/ could be the element needed to
explain the population of massive young stars observed today (Hansen \&
Milosavljevi\'c \cite{hansen}).

 To precise the spectral type of the components, better constrain the mass
 of the IMBH, spectroscopy in the 1-5\micron\/ range of all the individual
 sources within IRS~13E, at angular resolution as good as 0.1\arcsec, is
 required.  In the L and M bands it should confirm the expected featureless
 spectrum, except dust signatures, of the IRS~13E3 and 13E5 objects. These
 studies will need near-infrared 3-D spectrometers behind an AO system on a
 8-m telescope, like SINFONI (Mengel \etal \cite{mengel} and AMBER behind
 VLTI (Petrov \etal \cite{petrov}). Deeper AO imaging, as already obtained
 with NAOS/CONICA north of the IRS~13E center (Eckart \etal
 \cite{eckart2}), could make it possible to detect more members of the
 cluster.  Proper motions of the fainter members would help to confirm
 which of the individual sources are kinematically bound together.
 Theoretical work is needed to confirm whether IRS~13E can be the remnant
 of a massive cluster. More generally, to address the problem of recent
 star formation in the vicinity of \SgrA\/ a full census of the hot star
 population, illustrated in this paper on a small field, remains to be
 completed. It can be done by deep AO imaging to the condition to acquire
 data down to 1\micron. Up to now, such data, which appeared essential as
 shown in the current analysis, have only been possible with NICMOS on
 HST. Search for optical counterparts of the star-like X-ray sources
 detected by Chandra in the central parsec is another objective.

\begin{acknowledgements}

 We gratefully acknowledge helpful discussions with R. Coker which
stimulated the close examination of the high-resolution data available on
IRS~13. We want to warmly thank Y. Cl\'enet (Meudon Observatory) who made
available to us his L band AO data. Thanks to the organizing committee led
by Tom Geballe, the GC02 Workshop in Kona (Nov. 2002) was an ideal place
to improve various issues raised in this paper. Special thanks are also due
to T. Ott who sent us the proper motions of the sources of the IRS 13
field, prior to publication.

\end{acknowledgements}

\end{document}